\documentstyle[11pt,paspconf,epsf]{article}

\begin{document}

\title{Infrared interferometry of circumstellar envelopes}

\author{John D. Monnier}
\affil{Physics Dept, University of California,
    Berkeley, CA 94720-7450, USA}

\begin{abstract}
This paper will review the technical progress of interferometric
infrared observing techniques from the first 2-element interferometer
25 years ago to the 3+ element arrays now coming into service.  To
date, only the Infrared Spatial Interferometer (ISI) has published
separate-element interferometric data on circumstellar dust shells in
the infrared and many of these scientific results will be discussed.
Speckle interferometry has also evolved significantly over
the last few decades as slit-scanning techniques over single-pixel
detectors have largely been replaced by fast-readout of large format
detector arrays.  Important near-infrared and mid-infrared results
derived from speckle data will also be reviewed.  

Until recently, two-dimensional information about circumstellar dust
distributions has been sorely lacking, hence detections of dust shell
asymmetries have been difficult and uncertain.  New speckle
observations using modern, 10-m class telescopes have yielded
surprising results, demonstrating the importance of accurate closure
phase information in interpreting interferometric data.  These
discoveries hopefully precursor those to be made from closure-phase
imaging with the new generation of separate-element, interferometric
arrays.
\end{abstract}




\section{Introduction}
Infrared interferometry arises from the desire to image the heavens in
ever increasing spatial resolution.  Other techniques exist with the
same goal in mind, but the limited space afforded this review
necessarily restricts the scope to exclude lunar occultation (e.g.,
Ragland, Chandrasekhar \& Ashok 1997; Kaufl et al.\ 1998) and adaptive
optics work (e.g., Cruzalebes et al.\ 1998; Monnier et al.\ 1999).
This review will also ignore the rapidly increasing interferometric
successes at observing the stellar surfaces themselves (Dyck et al.\
1996; Wilson, Dhillon \& Haniff 1997; Burns et al.\ 1998; Van Belle et
al.\ 1998; Tuthill, Monnier \& Danchi 1999; Young et al.\ 1999), but
rather will concentrate on the circumstellar envelopes of late-type stars.

The objects under scrutiny here are mass-losing asymptotic giant
branch (AGB) stars; many of these stars are Mira variables, both with
oxygen- and carbon-rich shells.  This volume of conference papers is
devoted to the study of these objects and I invite you to look through
the other review articles in order to gain a greater understanding of
these interesting astrophysical laboratories.  That said, what do
interferometers hope to measure and then hopefully learn about the
astrophysics of the AGB?  Interferometrists strive to measure the most
detailed information possible concerning the
morphologies of circumstellar dust shells.  This requires observations
at various wavelengths as well as at various times.  These
measurements then can be used to directly test theories of dust
production and mass-loss mechanisms in general, which are important in
guiding our understanding of stellar evolution along the AGB.  When
coupled with spectral line observations, high angular resolution
studies can directly measure gas densities, velocities, and
temperatures around these stars, as well as probe the chemistry of
molecular formation.

\section{Introduction to Interferometry}
Although images, or maps, can be made from interferometry data (a
process known as aperture synthesis), an interferometer actually
measures a quantity called the {\em visibility}.  When light beams from two
separate telescopes are brought together to produce a fringe signal,
the visibility is proportional to the strength of the detected
fringes.  The visibility, which is a function of the projected
telescope separation (or baseline), can be shown to be equivalent to
one of the Fourier component amplitudes of the object brightness distribution.
Nice discussions of interferometry
can be found in Thompson, Moran \& Swenson (1986) and Born \& Wolf (1964).

For reconstructing reliable images of complex extended objects, one
also requires accurate estimates of the Fourier phases, not just the
amplitudes.  While atmospheric turbulence scrambles the phase of
individual Fourier components, a quantity called the {\em closure
phase} retains its integrity.  By summing the fringe phases for
baselines in a closed triangle, the closure phase is formed and allows
a majority of the intrinsic phase information to be recovered for
interferometric arrays with three or more telescopes (e.g.,
Jennison 1958; Readhead et al.\ 1988).  This quantity is critically
important for accurate reconstructions of highly asymmetric structures
in images.

Since any image can be alternatively represented by its Fourier
components, the collection of all ``interesting'' components allows
the interferometric data to be inverted, thus producing an image.
The Caltech VLBI and the VLBMEM packages can be used
to estimate the original object brightness distribution using Fourier
inversion algorithms such as CLEAN (with self-calibration) and the
Maximum Entropy Method (e.g., H\"{o}gbom 1974; Cornwell \& Wilkinson
1981; Sivia 1987; Haniff et al.\ 1989).  Speckle interferometry data
(more on this later) is often reduced using a bispectral analysis code
which performs a direct inversion from the estimated Fourier
amplitudes and phases.  However, modifications of this technique lead
to more robust reconstructions in the presence of noise (e.g., Hofman
\& Weigelt 1993).

\section{First Generation of Infrared Michelson Interferometers}
The first interferometric observations of circumstellar dust shells
were performed by the Arizona group, appearing in McCarthy \& Low
(1975).  The interferometer setup was mounted at the Cassegrain focus
of a large telescope, interfering together beams from two subapertures
onto a single-pixel InSb detector (later upgraded to a Ge
bolometer).  Additional observations were presented in McCarthy, Low,
\& Howell (1977), McCarthy, Howell, \& Low (1978, 1980), and McCarthy
(1979).  These measurements were sensitive in the mid-infrared, 
usually between 5\,$\mu$m to 12.5\,$\mu$m.

Meanwhile C. H. Townes was leading the U.C. Berkeley group in building
a separate-element heterodyne interferometer.  Infrared light incident
on each of the Auxiliary McMath Solar Telescopes at Kitt Peak (5.5\,m
separation) were combined with a CO$_2$ laser operating near 10
microns to produce a down-converted RF signal from a Ge:Cu detector
(later HgCdTe).  The RF signals (original bandwidth $\approx
\pm$1200\,MHz) from the two telescopes were sent through a delay line
and subsequently correlated, acting much like a typical radio or
mm interferometer.  This system detected fringes on the limb of
Mercury as early as 1974 (Johnson 1975), and observations of
circumstellar envelopes soon followed (Sutton et al. 1977, 1978, 1978).

These two groups observed many of the same stars and reached the same
general conclusions.  A diversity of dust shell types were detected,
both large and small (in terms of stellar size).  Both groups measured
changes in the dust shell around the Mira variable $o$~Cet, and attributed
these to the large amplitude pulsation.  These first set of
measurements also indicated asymmetries for some of the thickest (i.e. most 
resolved) dust shells (McCarthy 1979).

\section{One-dimensional Speckle Interferometry}
The promise of speckle interferometry was first realized by Antoine
Labeyrie (Labeyrie 1970).  In the original formulation, short
exposures of an astrophysical object are made to freeze the
``speckling'' induced by the turbulent atmosphere.  The amount of
high-resolution structure in the speckle pattern, as quantified by its
power spectrum, is a measure of two things: 1) the quality of the
atmospheric seeing, and 2) the high resolution structure in the object
of interest.  Observing a nearby point-source star allows the
calibration of the seeing contribution and thus the extraction of
interferometric visibility measurements out to the diffraction limit of
the telescope. It was not until later that the Fourier phases were
estimated from such data (e.g., Knox \& Thompson 1974; Weigelt 1977).

Although speckle interferometry was applied to measurements of
visible star diameters and binary star orbits in the early- to mid-1970's, it
was not until 1979 that the first infrared application to AGB
circumstellar envelopes emerged.  A group from France published a
number of papers on the subject, resolving a few dust shells for the
first time in the near-infrared (Foy et al.\ 1979; Sibille, Chelli \&
Lena 1979; Mariotti et al.\ 1983).  Because of the lack of
two-dimensional array cameras, a single pixel detector (typically
InSb) was used to sample the speckle cloud.  In most setups, a slit
was scanned across the speckled image (or the image was passed over
the slit).  Power spectra for all scans were averaged together to
estimate the visibility, after calibrating the mean atmospheric (and
telescope) transfer function.

While many groups made speckle observations in the 1980's, I want to
discuss the few representing significant advances.  Dyck et al. (1984)
contains the largest sample of resolved dust shells to date, observing
16 late-type stars and proto-planetary nebula (PPN) between
1.65\,$\mu$m and 4.8\,$\mu$m.  This work was able to confirm, in a
general sense, that dust condenses a few stellar radii from the
stellar surface and the existence of dust shell asymmetries for a
number of the most resolved sources.

Another significant advance occurring was the use of radiative transfer
models for interpreting visibility curves.  Because spatial resolution
is often inadequate to unambiguously interpret the data, radiative
transfer codes allow the inner radius of dust formation to be
estimated and can utilize multi-wavelength data in a natural way.
Two illustrations can be found in Ridgway et al. (1986) and Ridgway \&
Keady (1988), which estimated the dust shell optical depths for
IRC~+10216, NML Cyg, and IRC~+10420.

A number of other firsts occurred during this very productive time.
Dyck, Beckwith, \& Zuckerman (1983) published the first (and only?)
infrared speckle observations on spectral lines, five fundamental
vibration-rotation lines of CO around IRC~+10216.  Phase retrieval
methods were first applied to reconstructing 
1-D profiles of asymmetric dust shells
(Chelli, Perrier \& Biraud 1983; Leinert \& Haas 1989; Navarro,
Benitez \& Fuentes 1990), and speckle techniques were applied in the
mid-infrared (Cobb \& Fix 1987; Benson, Turner \& Dyck 1989).  Another
Dyck paper, Dyck et al. (1991), compared the existing IRC~+10216
speckle data to show that the circumstellar emission changed
significantly during the 1980's.  There were a number of other
interesting near-infrared (e.g., Dainty et al.\ 1985; Dyck et al.\
1987; Mariotti et al.\ 1992; Lopez et al.\ 1993) and mid-infrared
speckle results (Fix \& Cobb 1988; Dyck \& Benson 1992) published
during this time.

\section{The Infrared Spatial Interferometer}
After the successful interferometry experiment at Kitt Peak, the UC
Berkeley group began constructing a pair of movable telescopes to be
placed on Mt.  Wilson, California, based on the same heterodyne
detection design (present bandwidth $\approx
\pm$2700\,MHz).  The Infrared Spatial Interferometer (ISI) published
its first results on IRC +10216 in 1990 (Danchi et al.\ 1990; 
Bester et al.\ 1991) 
and showed
the dust shell size changes as a function of the stellar pulsational phase.

The most significant single paper was a compilation of dust shell
sizes for 13 late-type stars (Danchi et al.\ 1994), which established two
classes of dusty stars, those with hot dust close to the star and those
with detached, relatively cool shells.  In addition, the variation of
inner radii indicated a episode emission time scale of 20-50 years.
The ISI has monitored a number of AGB stars for a 10-year period now,
discovering complex shell geometries and time variation around multiple
sources (Greenhill et al.\ 1995; Bester et al.\ 1996; 
Monnier et al.\ 1997;
Hale et al.\ 1997; Lopez et al.\ 1997).  High spatial resolution observations
with $\frac{\lambda}{\Delta \lambda} \approx 10^5$ presently underway
should reveal the location of NH$_3$ and SiH$_4$ molecules, and help
explain their anomalously high observed abundances around AGB stars
(e.g., Betz, McLaren \& Spears 1979; Goldhaber 1988).

\section{Two-dimensional Speckle Interferometry}
Two major technical advances in the 1990's have ushered in a
renaissance in speckle observations.  The development of fast-readout,
infrared array cameras allows the entire speckle cloud to be recorded
and thousands of images to be collected in a short period of time.
Secondly, a new generation of large aperture telescopes has come into
service, allowing 2-3 times higher spatial resolution.

Most present results are coming from two groups using different techniques and
telescopes, although other groups are making significant
contributions as well (e.g., Haniff \& Buscher 1998; 
Buscher, Haniff \& Oudmaijer 1998).  A team
headed by G. Weigelt has been using so-called ``speckle masking,'' a
modified bispectral analysis, to analyze speckle data collected at the
6m Special Astrophysical Observatory telescope and the 3.5m+ ESO
telescopes on La Silla, Chile.  Very nice results have recently been
reported (Osterbart et al. 1997; Weigelt et al. 1998; Osterbart et al.\ 1998;
Irrgang et al.\ 1998; Weigelt et al.\ 1999).
P. Tuthill leads the other group,
which uses aperture masking techniques on the Keck-I telescope
with maximum baselines of 11m.  Results can be found in this volume
(Tuthill et al.\ 1999) and in other publications
(Tuthill, Monnier \& Danchi 1998; Monnier et al.\ 1999).

Both groups have succeeded in constructing very detailed images at the
telescope diffraction limit.  The resolution achieved equals about one
stellar radii for the closest objects (e.g., IRC +10216), allowing the
dust formation process to be studied in detail.  Three different
groups presented images of IRC +10216 at 2.2\,$\micron$, and all
images showed the same asymmetric and clumpy dust shell (Weigelt et
al.\ 1998; Haniff \& Buscher 1998; Tuthill et al.\ 1999).  In
addition, proper motions of the dust clouds have been detected, which
promise to yield important dynamical and three-dimensional
information on the dust shell geometry.  These observations are
showing that mass-loss processes can be very inhomogeneous in the
inner dust shell regions, for reasons not entirely clear.

\section{Conclusions}
I've summarized the progress of infrared interferometry on
circumstellar envelopes in Figure~1 and Table~1.  Figure~1 shows the
spatial resolution and wavelength coverage attainable with present
instruments, while Table~1 lists all dust shells resolved by papers
referenced here (and is a fairly complete list of published results).
This list will surely grow by a dozen or so in the coming years just
from analyses of existing data, especially from the speckle imaging
groups.  While many measurements do exist, a lack of coordinated
observations at multiple wavelengths has hindered interpretation of
these data because of the time variable nature of these sources.
Also, combining speckle observations with polarization and spectral
line work can make unique and valuable contributions to the study of
AGB stars, but have only rarely been performed.  In addition,
photometric monitoring of these sources in the infrared can be very
valuable when trying to understand temporal changes in the dust shell
properties, but is also rarely done, despite their high flux levels.

\begin{figure}
\plotone{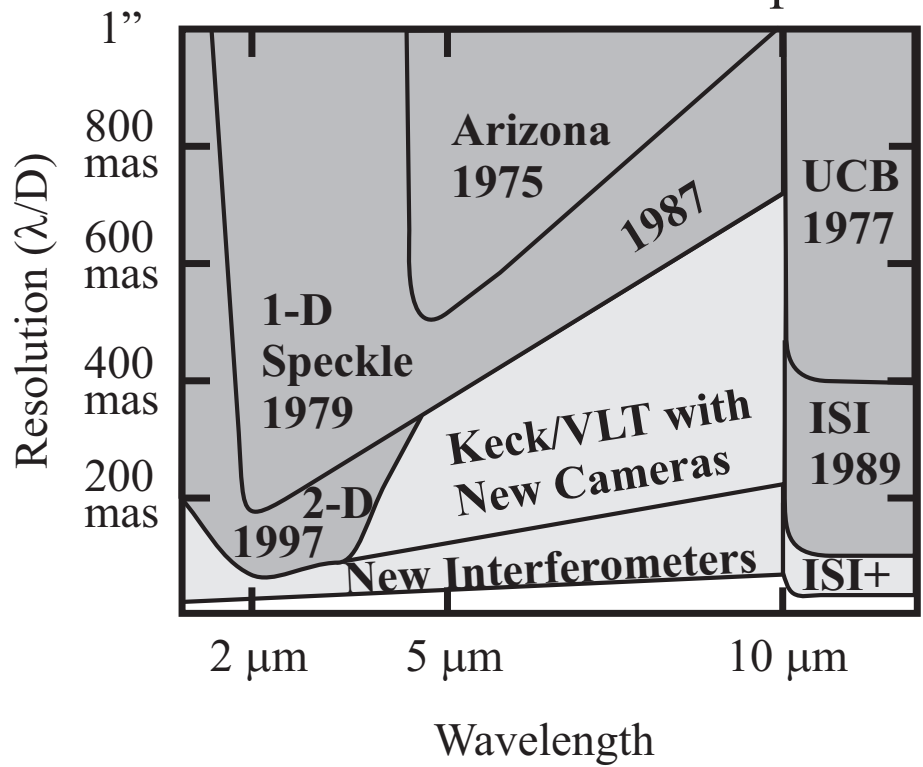}
\caption{Spatial resolution vs. wavelength for past, present, and future
interferometric observations.  Parameter space to be covered in the
next few years is shaded in light grey.}
\label{final}
\end{figure}

The inhomogeneous, asymmetric and time-variable dust shells observed
around many stars indicate that mass-loss on the AGB may not be
understood quite as well as previously thought.  More theoretical work
and modeling (two- or even three-dimensional radiative transfer) are
needed to make full use of the new high resolution data.
Understanding of these processes should continue to improve as a
number of infrared interferometer arrays are being constructed or
upgraded (IOTA, COAST, PTI, CHARA, Keck, VLTI, MIRA, ISI+, more),
leading to spatial resolutions (at some wavelengths) of 
less than 1~mas in the near future.

\begin{table}
\caption{Circumstellar Shells of Late-type Stars Resolved by Infrared Interferometry \label{table1}}
\begin{center}\scriptsize
\begin{tabular}{lccl}
\tableline
Star Name	& Type & Wavelengths ($\mu$m)	& Other Comments \\
\tableline
NML Cyg		& Red Supergiant & 2.2, 2.4, 3.5,   & Asymmetric \\
		&		 & 3.6, 3.8, 4.8,    & \\ 
		& 		 & 5.0, 7.9, 8.7,   & \\
 		&		 & 10, 10.4, 11.4, 12.6  &  \\
Alp Ori		& Red Supergiant & 7.9, 8.7, 10,    & Time variable \\
		&		 & 10.2, 10.4, 11.1, 11.4& \\
VY CMa		& Red Supergiant & 1.65, 2.2, 3.1, & Asymmetric \\
		&		 & 3.8, 4.8, 8.3,  & 2-D images \\
  		& 		 & 10.2, 11.1 &  \\
S Per		& Red Supergiant& 7.9, 8.7, 10, & \\
		&		& 10.4, 11.4, 12.6 & \\
Alp Sco		& Red Supergiant & 10.0, 11.1 &  Asymmetric, 1-D profile\\
VX Sgr		& Red Supergiant & 2.2, 3.8, 11.15 & \\
Mu Cep		& Red Supergiant& 10 & \\
Alp Her         & Red Supergiant & 11.15 \\
IRC +10216 	& Carbon Star    & 2.2, 3.1, 3.5,   & Asymmetric, Time variable, \\
 		&		 & 8.3, 10, 10.2,  &  Polarization data, 1-D profiles, \\
		&		 & 11.1, 12.5		       &  Spectral Line Work (CO, NH$_3$, SiH$_4$), \\
		&		 &	 & 2-D images, \\
		&		 &	 & Proper motions detected \\
CIT 6 		& Carbon Star	 & 2.2, 3.1, 3.8,  & Asymmetric, \\
		& 		 & 4.8, 10, 11.5 & 2-D images \\
IRAS 15194-5115  & Carbon Star   & 2.2, 3.6 & \\  
CIT 5		& Carbon Star	 & 4.8  & \\

Chi Cyg		& Mira           & 2.2, 4.6, 4.8, & Asymmetric \\
		&		 & 10, 11.15  & \\
IK Tau		& Mira 		 & 3.8, 11.15 & Time variable\\
Omi Cet		& Mira           & 10.2 & Asymmetric, Time Variable \\
R Leo		& Mira           & 11.1 &  \\
R Cas 		& Mira		& 7.9, 8.7, 10, & \\
		& 		& 10.4, 11.4, 12.6 & \\
CIT 3		& Mira		& 1.65, 2.2, 11.15 & Asymmetric \\
AFGL 2290	& Mira		& 2.2 & Asymmetric \\ 
T Cas		& Mira  	& 10 & \\
U Her		& Mira		& 10 & \\
U Ori 		& Mira 		& 11.15 & \\
W Aql		& Mira		& 11.15 & \\
R Aqr		& Mira		& 11.15 & \\

BC Cyg		& Mira? 	& 7.9, 8.7, 10, & \\
		& 		& 10.4, 11.4, 12.6 & \\
BI Cyg		& Mira?		& 7.9, 8.7, 10, & \\
		&		& 10.4, 11.4, 12.6 & \\

RX Boo		& Semiregular	& 7.9, 8.7, 10, & \\
		&		& 10.4, 11.4, 12.6 & \\
SW Vir		& Semiregular    & 2.4, 10 & \\
X Her		& Semiregular   & 10 & \\
OH 26.5+0.6	& OH/IR		 & 3.8, 4.6, 4.8, & Asymmetric \\
		&		 & 5, 10  &  \\
OH 354.9-0.5 	& OH/IR		& 10 & Asymmetric \\
OH 1.1-0.8	& OH/IR		& 10 & Asymmetric \\

Red Rectangle  	& PPN		 & 1.25, 1.6, 2.2, & Asymmetric, 1-D Profile, \\
		&		 & 3.1, 3.3, 3.8,  & Polarization data, \\
		&		 & 4.8    & 2-D images \\
Calabesh Nebula & PPN 		 & 3.8 & Asymmetric \\
AFGL 618        & PPN            & 3.8 & \\
M1-9 		& PPN		 & 3.8, 4.8 & Asymmetric \\
M2-9		& PPN 		 & 3.8 & Asymmetric \\

IRC +10420 	& F8-G0 I        & 2.2, 3.5, 3.8, & Asymmetric \\
		&		 & 4.8, 5, 10 &  \\
Eta Car	        & LBV 	 	 & 4.6 & 1-D profiles, 2-D images \\

\tableline
\end{tabular}
\end{center}



\end{table}

\acknowledgments

I am thankful to A. P. Zemgulys for enlightening discussions.


\begin{question}{Audience Member}
What is the faintest object that can be imaged using speckle
interferometric techniques on large telescopes?
\end{question}
\begin{answer}{J. D. Monnier}
The aperture masking experiment at Keck has succeeded in obtaining 
good visibility and Fourier phase information down 
to K-band magnitude of 5.  We have not pushed dimmer than this yet,
but believe that standard speckle techniques will need to be used to go much
dimmer.
\end{answer}
\begin{answer}{G. Weigelt}
We have gone down to K-band magnitude 9 for our observations of NGC~4151
using ``speckle masking'' [not to be confused with ``aperture masking''].
\end{answer}

\begin{question}{Audience Member}
What is your interpretation of the complicated dust emission seen around
IRC~+10216?
\end{question}
\begin{answer}{J. D. Monnier}
I would say the interpretation is certainly still controversial, but
the Keck observations of a compact core at H, K, and 3.1\,$\mu$m with
an angular diameter of about 50~mas, suggest the bright southern
component is indeed the central carbon star itself.  This should be
decided definitively in the next few years as we continue to track the
proper motions of the various dust clouds in the outflow [see 
Tuthill et al. (1999) in this volume].
\end{answer}

%
%


\end{document}